\documentclass[preprint,tightenlines,aps,floats,nofootinbib,showpacs]{revtex4}

\usepackage{epsf}
\usepackage{epsfig}
\usepackage{graphicx}
\usepackage[dvips]{color}

\newcommand{\notE}{\ \hbox{{$E$}\kern-.60em\hbox{/}}}
\newcommand{\notp}{\ \hbox{{$p$}\kern-.43em\hbox{/}}}
\def\D0{\mbox{D\O}}

\newcommand{\eps}{\epsilon}

\includegraphics{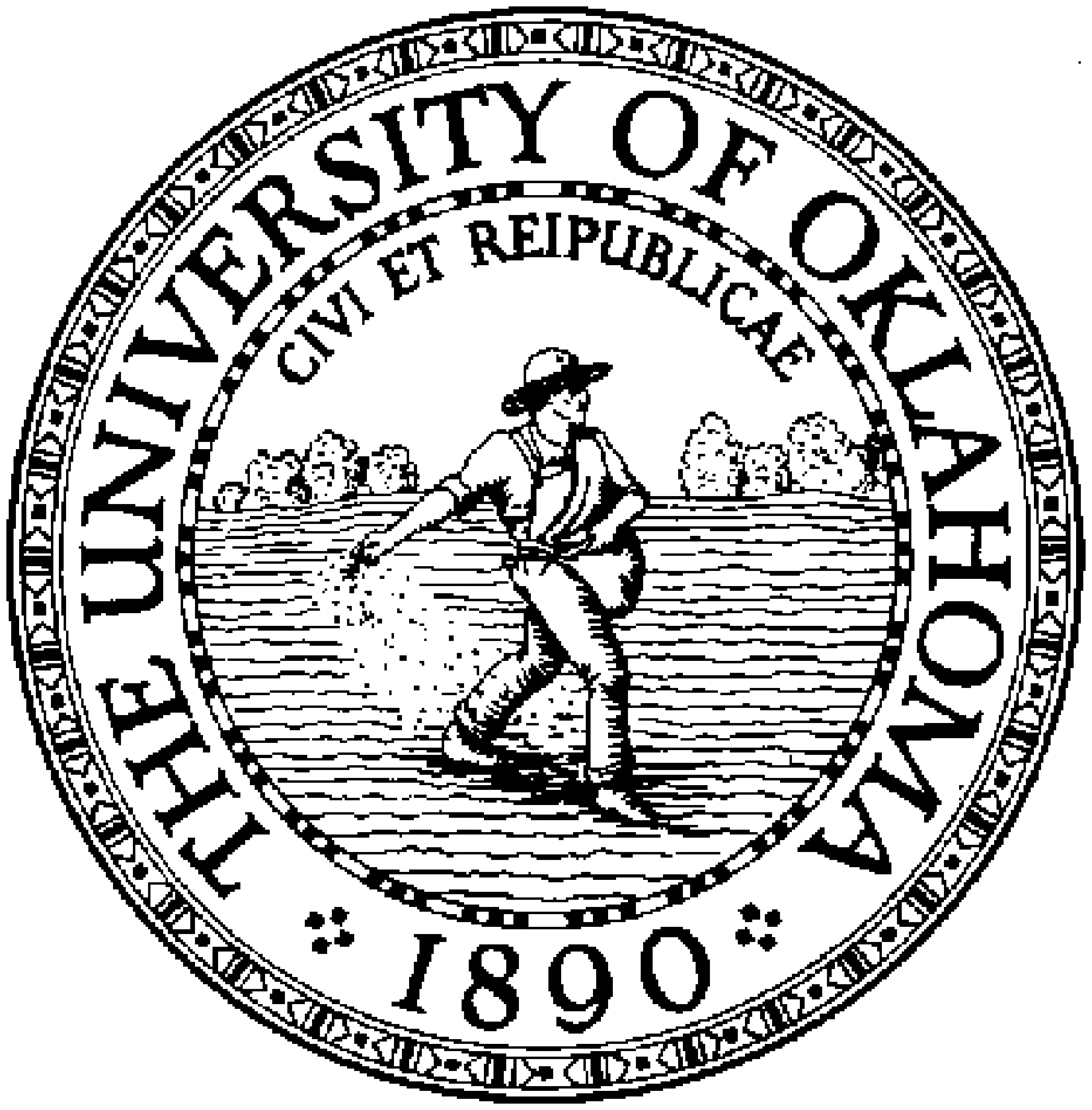}

\preprint{\font\fortssbx=cmssbx10 scaled \magstep2
\hbox to \hsize{
\hskip1.2in 
\hbox{\fortssbx The University of Oklahoma}
\hskip0.2in $\vcenter{
                      \hbox{\bf OKHEP-07-02}
                      \hbox{\bf SLAC-PUB-12963}
                      \hbox{\bf UH-511-1102-07}
                      \hbox{\bf arXiv:0711.0232 [hep-ph]}
                      \hbox{November 2007}}$ }
}

\begin{document}

 
\title{\vspace*{0.7in}
Discovering the Higgs Bosons of Minimal Supersymmetry \\ 
with Tau Leptons and a Bottom Quark}
 
\author{
Chung Kao$^{a,b}$, Duane A. Dicus$^c$, Rahul Malhotra$^d$ and Yili Wang$^a$}

\affiliation{
$^a$Department of Physics and Astronomy, University of Oklahoma, 
Norman, OK 73019, USA \\
$^b$Stanford Linear Accelerator Center, 
2575 Sand Hill Road, Menlo Park, CA 94025, USA \\
$^c$Center for Particles and Fields, University of Texas, 
Austin, TX 78712, USA \\
$^d$Department of Physics and Astronomy, University of Hawaii,
Honolulu, HI 96822, USA
\vspace*{.5in}}

\date{\today}

\thispagestyle{empty}

\begin{abstract}

We investigate the prospects for the discovery 
at the CERN Large Hadron Collider or at the Fermilab Tevatron
of neutral Higgs bosons through the channel where the Higgs are
produced together with a single bottom quark 
and the Higgs decays into a pair of tau leptons, 
$bg \to b\phi^0 \to b\tau^+\tau^-, \phi^0 = h^0, H^0, A^0$.
We work within the framework of the minimal supersymmetric model.
The dominant physics background from the production of $b\tau^+\tau^-$, 
$j\tau^+\tau^-$ ($j = g, u, d, s, c$), $b\bar{b}W^+W^-$, $W+2j$ 
and $Wbj$ is calculated with realistic acceptance cuts and efficiencies.
Promising results are found for the CP-odd pseudoscalar ($A^0$) 
and the heavier CP-even scalar ($H^0$) Higgs bosons with masses 
up to one TeV.

\end{abstract}


\maketitle

\newpage

\section{Introduction}

In the minimal supersymmetric standard model (MSSM) \cite{MSSM}, 
the Higgs sector has 
two doublets,
$\phi_1$ and $\phi_2$, which couple to fermions with weak isospin 
$t_3 = -1/2$ and $t_3 = +1/2$ respectively \cite{Guide}. 
After spontaneous symmetry breaking, there remain five physical Higgs bosons:
a pair of singly charged Higgs bosons $H^{\pm}$,
two neutral CP-even scalars $H^0$ (heavier) and $h^0$ (lighter),
and a neutral CP-odd pseudoscalar $A^0$.
The Higgs potential is constrained by supersymmetry 
such that all tree-level Higgs boson masses and couplings 
are determined by just two independent parameters,  
commonly chosen to be the mass of the CP-odd pseudoscalar ($M_A$) 
and the ratio of vacuum expectation values of the neutral Higgs fields 
($\tan\beta \equiv v_2/v_1$). 

At the CERN Large Hadron Collider (LHC), 
gluon fusion ($gg \to \phi, \phi = h^0, H^0$, or $A^0$) is the major 
source of neutral Higgs bosons in the MSSM for $\tan\beta$ less than about 5.
If $\tan\beta$ is larger than 7, neutral Higgs bosons are dominantly 
produced from bottom quark fusion $b\bar{b} \to \phi$ 
\cite{Dicus1,Dicus2,Balazs,Maltoni,Harlander}. 
Since the Yukawa couplings of $\phi b\bar{b}$ are enhanced by $1/\cos\beta$,
the production rate of neutral Higgs bosons, especially the $A^0$ 
or the $H^0$, is enhanced at large $\tan\beta$.

Recently, it has been suggested that the search for a Higgs boson 
produced along with a single bottom quark with large transverse 
momentum ($p_T$), where the leading order subprocess is $bg \to b\phi$ 
\cite{Choudhury,Huang,Scott,Cao,Dawson:2007ur}, could be more promising 
than the production of a Higgs boson associated with two high $p_T$ 
bottom quarks \cite{Scott} where the leading order subprocess is 
$gg \to b\bar{b}\phi$ \cite{Dicus1,hbbmm,Plumper,Dittmaier,Dawson}.
This has already been demonstrated to be the case for the $\mu^+ \mu^-$ 
decay mode of the Higgs bosons \cite{hbmm}. 
For a large value of $\tan\beta$, 
the $\tau^+\tau^-$ decay mode \cite{Kunszt,Richter-Was}
is also a promising discovery channel for the $A^0$ and the $H^0$ 
in the MSSM because the branching fraction for Higgs 
decay into tau leptons is greater by a factor of 
$(m_{\tau}/m_{\mu})^2 \sim 286$. The downside is that unlike muons, 
tau leptons can only be observed indirectly 
via their hadronic or leptonic decay products.

 
In this article, we present the prospects of discovering the MSSM neutral 
Higgs bosons produced with a bottom quark via Higgs decay into tau pairs. 
We calculate the Higgs signal and the dominant Standard Model (SM) 
backgrounds with realistic cuts and efficiencies and evaluate the 
$5\sigma$ discovery contour in the $(M_A,\tan\beta)$ plane for the LHC and for the Tevatron.

\section{The production cross sections and branching fractions}

We calculate the cross section at the LHC for $pp \to b \phi +X$ 
and at the Tevatron for $p\bar{p} \to b \phi +X$ 
($\phi = H^0, h^0, A^0$) via $bg \to b \phi$ 
with the parton distribution functions of CTEQ6L1 \cite{CTEQ6}. 
The factorization scale is chosen to be $M_{\phi}/4$ \cite{Maltoni,Plehn}.
In this article, unless explicitly specified,
$b$ represents a bottom quark ($b$) or a bottom anti-quark ($\bar{b}$). 
The bottom quark mass in the $\phi b\bar{b}$ Yukawa coupling 
is chosen to be the next-to-leading order (NLO) running mass
at the renormalization scale $\mu_R$, $m_b(\mu_R)$ \cite{bmass}, 
and it is calculated with $m_b({\rm pole}) = 4.7$ GeV and NLO evolution 
of the strong coupling \cite{alphas}. 
We have also taken the renormalization scale 
for the production processes to be $M_{\phi}/4$, which 
effectively reproduces the effects of next-to-leading order \cite{Scott}. 
Therefore, we take the $K$ factor to be one for the Higgs signal. 

The cross section for $pp \to b \phi \to b \tau^+\tau^- +X$ can be 
thought of as the Higgs production cross section $\sigma(pp \to b \phi +X)$ 
multiplied by the branching fraction of the Higgs decay into tau pairs
$B(\phi \to \tau^+\tau^-)$. When the $b\bar{b}$ mode dominates Higgs decays, 
the branching fraction of $\phi \to \tau^+\tau^-$ is about 
$m_\tau^2/(3 m_b^2(M_\phi)+m_\tau^2)$ where $m_b(M_\phi)$, 
the running mass at the scale $M_\phi$, is used in the decay rates.
This results in a branching fraction for $A^0 \to \tau^+\tau^-$ 
of $\sim 0.1$ for $M_A = 100$ GeV.
Thus for $\tan\beta \agt 10$ and $M_A \agt 125$ GeV, the cross section 
of $bA^0$ or that of $bH^0$ is enhanced by approximately $\tan^2\beta$ 
and the branching fraction of Higgs decay to tau pair is close to $10\%$.

\section{Tau Decay and Identification}

Tau leptons can decay either purely leptonically, 
$\tau^- \rightarrow \ell^- \bar{\nu}_\ell \nu_\tau$, with a branching ratio of 
around $18\%$ for each mode $l = e,\mu$, or they can decay into 
low-multiplicity hadronic states and a $\nu_\tau$ with a branching ratio 
$\simeq 64\%$ \cite{PDG}. 
Therefore, for a $\tau^+ \tau^-$ pair, the most likely scenario is one 
decaying leptonically and the other hadronically, which has a 
combined branching ratio of $46\%$. Also, the presence of an isolated 
lepton in the final state is useful in triggering the event 
and reducing backgrounds. Hence, we use this 
``lepton + $\tau$-jet'' signature in our study. 

We model hadronic tau decays as the sum of 
two-body decays into $\pi\nu_\tau$, $\rho\nu_\tau$ and $a_1 \nu_\tau$ 
with branching ratios given in the literature \cite{PDG}. 
The tau is assumed to be 
energetic enough that all its decay products emerge in approximately the same 
direction as the tau itself. This manifests itself in the 
so-called ``collinear approximation'' which we use for both 
leptonic and hadronic decays. The approximation is confirmed to be accurate 
by comparison with an exact matrix element simulation for tau decay.

In Figure 1, we present the transverse momentum distribution 
($d\sigma/d{p_T}$) for the bottom quark ($b$), 
or the lepton ($\ell$) or the tau hadron ($j_\tau$) from tau decays, 
for the Higgs signal $pp \to bA^0 \to b\tau^+\tau^- \to b\ell j_\tau +X$. 
In addition, we show the $p_T$ distribution for
$b$, $\ell$, or $j_\tau$ from the SM background 
$bg \to b\tau^+\tau^-$ (Drell-Yan).
We have required $p_T(b) > 10$ GeV and $|\eta_b| < 2.5$.
The purpose of this figure is to show these cross sections
before any other cuts have been applied.

 
\begin{figure}[htb]
\centering\leavevmode
\epsfxsize=6in
\epsfbox{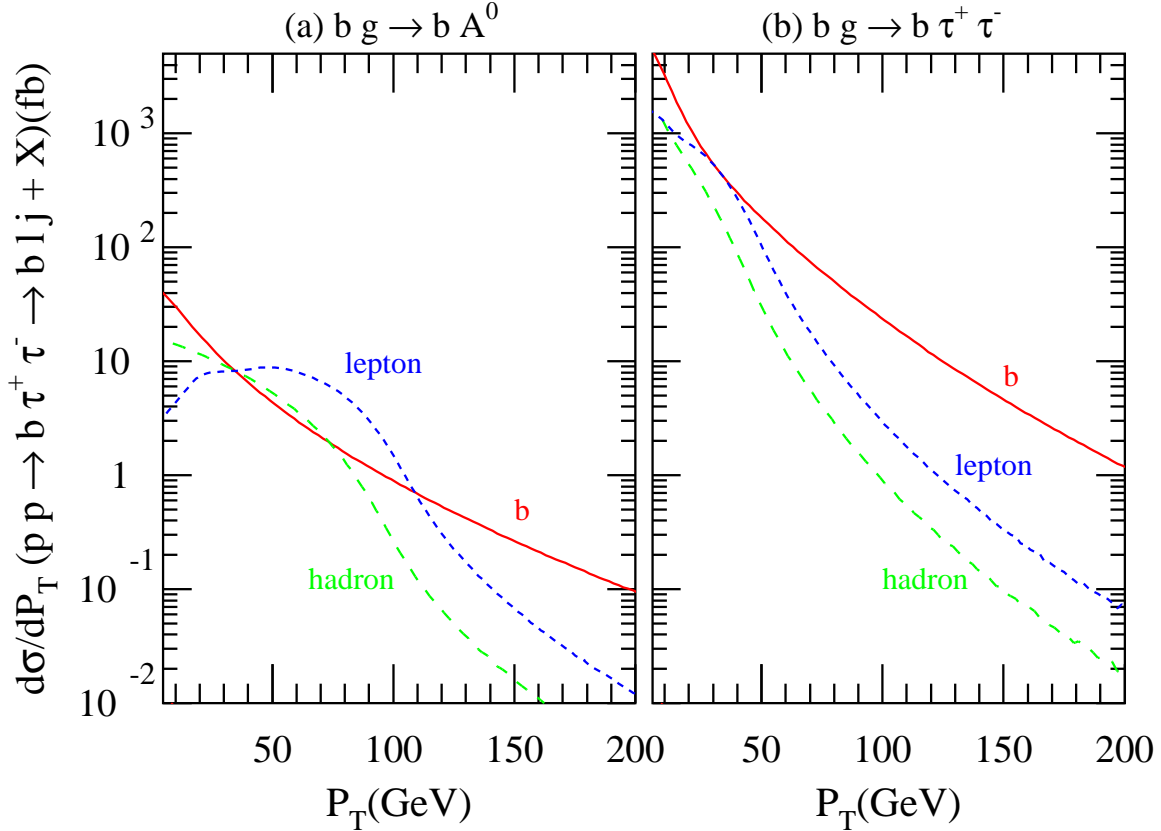}
\caption[]{
The transverse-momentum distribution for 
(a) the Higgs signal from $bg \to bA^0 \to b \tau^- \tau^+ \to b\ell j +X$ 
with $M_A = 200$ GeV and $\tan\beta = 10$ as well as 
for (b) the physics background from 
$bg \to b \tau^- \tau^+ \to b\ell j +X, j = \pi, \rho$, or $a_1$. 
In the three curves, $p_T$ refers to the transverse momentum of the
$b-$quark or the lepton or the tau-jet.
\label{fig:momentum}
}
\end{figure}

The ATLAS collaboration has studied identification efficiencies of 
$\tau$-jets in detail \cite{ATLAS}. 
Based on this we use an overall efficiency of $26\%$ over 1- and 
3-prong decays with a corresponding cut, 
$p_T(h) > 40$ GeV for the hadron 
$h = \pi, \rho, a_1$. This also corresponds to a mistag 
efficiency of $1/400$ for non-$\tau$ (i.e. QCD) jets. Rejection of jets from 
$b$ quarks is higher, with only 1 in 700 being mistagged as $\tau$s. 
The transverse momentum cut on the lepton from tau decay 
is weaker, with $p_T(\ell) > 20$ GeV. Both the hadron and lepton are 
required to be in the central rapidity region $|\eta| < 2.5$. 
The acceptance cuts as well as tagging and mistagging efficiencies for
the Fermilab Tevatron will be discussed in Section VII.

\section{Higgs Mass Reconstruction}

The Higgs mass can be reconstructed indirectly, using the collinear 
approximation for $\tau$ decay products and the missing transverse 
momentum 2-vector, ${\bf \notp_T}$. Taking $x_\ell, x_h$ to be the energy 
fractions carried away from the decays
by the lepton and hadron respectively, we have:
\begin{eqnarray}
(\frac{1}{x_\ell} - 1){\bf p_T^\ell} + (\frac{1}{x_h} - 1){\bf p_T^h} = 
{\bf \notp_T}
\end{eqnarray}
This yields two equations for $x_\ell$ and $x_h$ which can be solved to 
reconstruct the two original $\tau$ 4-momenta 
$p^{\mu}_\tau = p^{\mu}_\ell/x_\ell,$ $p^{\mu}_h/x_h$. 
Thus $M_\phi^2 = (p_\ell/x_\ell + p_h/x_h)^2$.
Physically we must have $0 < x_\ell,x_h < 1$, and this provides 
a further cut to reduce the background. 

Measurement errors in lepton and $\tau$-jet momenta as well as missing 
transverse momentum give rise to a spread in the reconstructed mass 
about the true value. Based on the ATLAS and the CDF specifications 
we model these effects by Gaussian smearing of momenta:
\begin{eqnarray}
\frac{\Delta E}{E} = \frac{0.50}{\sqrt{E}} \oplus 0.03
\end{eqnarray}
for jets (with individual terms added in quadrature) and
\begin{eqnarray}
\frac{\Delta E}{E} 
& = & \frac{0.25}{\sqrt{E}} \oplus 0.01 \;\; ({\rm LHC})  \\
\frac{\Delta E}{E} 
& = & \frac{0.15}{\sqrt{E}} \oplus 0.01 \;\; ({\rm Tevatron})  
\end{eqnarray}
for charged leptons.

We find that in more than $95\%$ of the cases, the reconstructed mass lies 
within $15\%$ of the actual mass. Therefore we apply a mass cut, requiring 
the reconstructed mass to lie in the mass window 
$M_\phi \pm \Delta M_{\tau\tau}$, where 
$\Delta M_{\tau\tau} = 0.15 M_\phi$ for an integrated luminosity ($L$)
of 30 fb$^{-1}$
and $\Delta M_{\tau\tau} = 0.20 M_\phi$ for $L =$ 300 fb$^{-1}$.
This cut is actually rather conservative because for larger Higgs masses, 
more than $90\%$ of the reconstructed masses are within $5-10\%$ of 
$M_\phi$. 
We note that improvements in the discovery potential will be possible 
by narrowing $\Delta M_{\tau\tau}$ if the $\tau$ pair mass resolution 
can be improved.

Figure 2 shows the invariant mass distribution of the tau pair
for the Higgs signal $pp \to bA^0 \to b\tau^+\tau^- +X$
via $bg \to bA^0$, and for the tau pair from the SM Drell-Yan background 
$bg \to b\tau^+\tau^-$. We have calculated the Higgs signal in two ways: 
(a) with the narrow width approximation 
\begin{eqnarray*}
 \sigma(pp \to bA^0 \to b\tau^+\tau^- +X) 
= \sigma(pp \to bA^0 +X) \times B( A^0 \to b\tau^+\tau^-)
\end{eqnarray*}
and 
(b) the full calculation 
$\sigma(pp \to bA^0 \to b\tau^+\tau^- +X)$ 
with a Breit-Wigner resonance
via $bg \to bA^0 \to b\tau^+\tau^-$. 
In this figure we have applied all acceptance cuts 
discussed in the next two sections
except the requirement on invariant mass. 
We note that with energy-momentum smearing, the cross section in
the narrow width approximation 
agrees very well with that evaluated for a Breit-Wigner resonance.

 
\begin{figure}[htb]
\centering\leavevmode
\epsfxsize=5in
\epsfbox{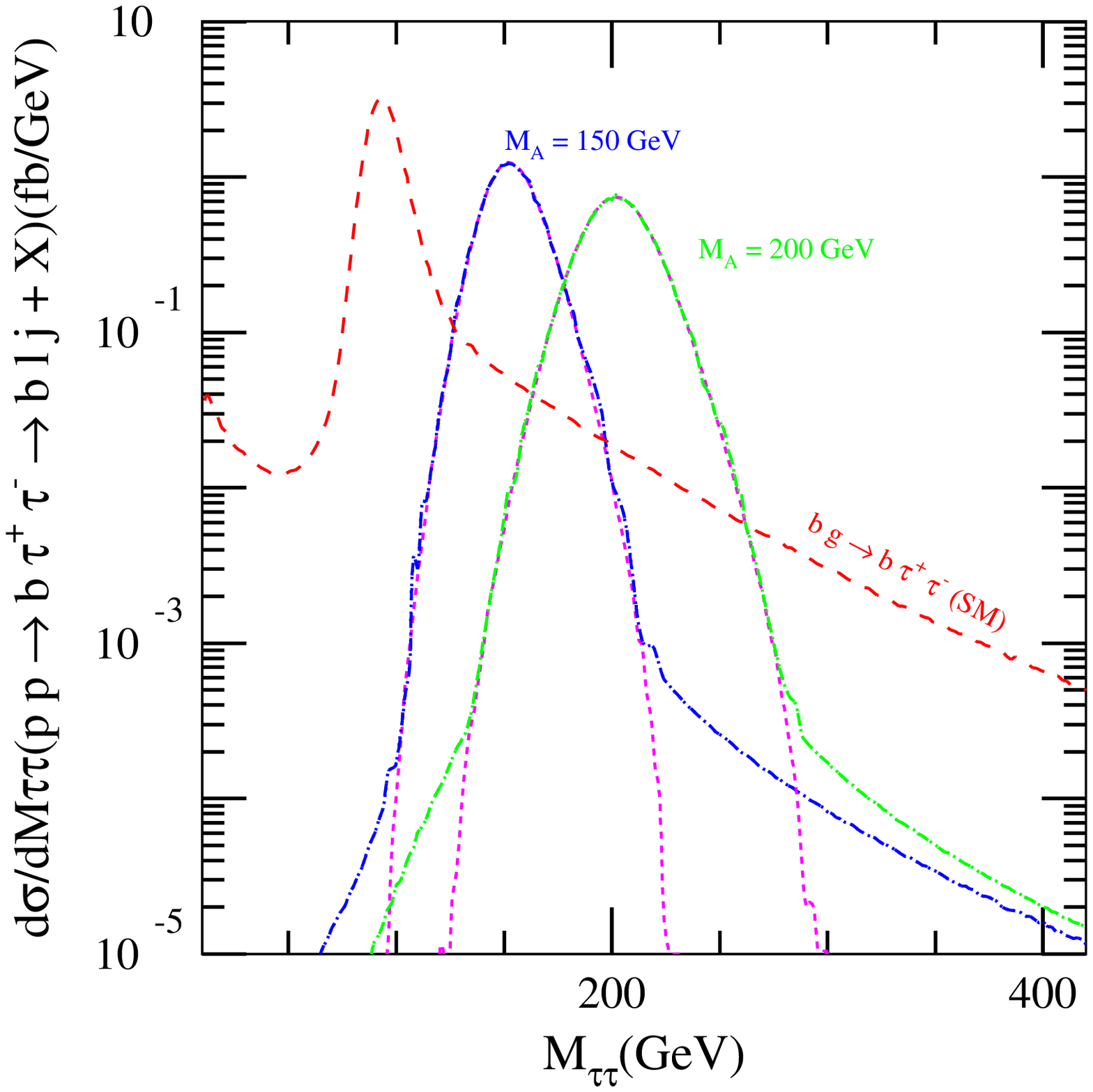}
\caption[]{
The invariant-mass distribution distribution, 
$d\sigma/dM_{\tau\tau}(pp \to b\tau^+\tau^- \to b\ell j_\tau +\notE_T +X)$, 
for the Higgs signal from 
$bg \to bA^0 \to \tau^- \tau^+ \to b\ell j_\tau +X$ 
with $M_A = 150$ GeV or 200 GeV and $\tan\beta = 10$ 
as well as for the physics background from 
$bg \to b \tau^- \tau^+ \to b\ell j_\tau +X, j_\tau = \pi, \rho$, or $a_1$ (dash).
We calculate the Higgs signal in the narrow width approximation (short dash)
and with a Breit-Wigner resonance from a Higgs propagator (dash-dot).
\label{fig:mass}
}
\end{figure}

\section{The Physics Background}

From the above discussion, the signal we are looking for is
$b$-jet ($b$) + lepton ($\ell$) + $\tau$-jet ($j$) + $\notE_T$ + $X$,
where $\notE_T =$ missing transverse energy $\simeq \notp_T =$ missing
transverse momentum.

The dominant physics backgrounds to this final state come from:
\begin{itemize}
\item[(i)] 
Drell-Yan processes: $pp \to jZ^*/\gamma^* + X \to j \tau^+\tau^- + X$, 
$j = u,d,s,c,b,g$. Approximately $60-70\%$ of the DY contribution arises 
from the subprocess $bg \to b\tau^+\tau^-$.
\item[(ii)]  
Top Production ($gg,q\bar{q} \to t\bar{t} \to b\bar{b}W^+W^-$):
This can contribute in several ways depending on how the  $W$s decay. 
In order of highest to lowest importance, the relevant channels are
the following.
(a) $b\tau\nu b\ell\nu$:
One $W$ decays into $\tau\nu_\tau$ with the $\tau$ decaying
hadronically while the other $W$ provides $\ell\nu_\ell$. 
(b) $b\tau\nu b\tau\nu$:
We can have both $W$'s decaying into $\tau\nu_\tau$ with one 
tau decaying leptonically and the other hadronically. 
(c) $b\ell\nu bjj$:
In this case we can have one $W$ decay leptonically 
while the other $W$ decays into jets ($W \to qq'$). 
We now have four possible jets in the final state
i.e., 2 $b$'s and $2j$ and one of them is tagged as a $b$ quark while 
one of the other is mistagged as a $\tau$-jet. 
(d) $b\ell\nu b\ell\nu$:
We can have both $W$s decay leptonically. Then we have two $b$ quarks 
in the final state, one of them is tagged as a $b-$jet while 
the others is mistagged as a $\tau$-jet. 
(e) $b\tau\nu bjj$:
Finally we can have one $W$ decay into $\tau\nu_\tau$ with the $\tau$ decaying
leptonically while the other $W$ decays into jets ($W \to qq'$). 
We now have four possible jets in the final state
i.e., 2 $b$'s and $2j$ and one of them is tagged as a $b$ quark while 
one of the others is mistagged as a $\tau$-jet. 
\item[(iii)]  
$tW$ Production ($bg \to tW \to bW^+W^-$):  
This is very similar to the top quark pair production just discussed. 
In order of decreasing importance, the relevant channels
are as follows.
(a) $b\tau\nu\ell\nu$:
One $W$ decays into $\tau\nu_\tau$ with the $\tau$ decaying
hadronically while the other $W$ provides $\ell\nu_\ell$. 
(b) $b\tau\nu\tau\nu$:
We can have both $W$'s decaying into $\tau\nu_\tau$ with one 
tau decaying leptonically and the other hadronically. 
(c) $b\ell\nu jj$:
In this case we can have one $W$ decay leptonically 
while the other $W$ decays into jets ($W \to qq'$). 
We now have one $b$ and $2j$ with 
one of the light quarks mistagged as a $\tau$-jet. 
(d) $b\tau\nu jj$:
Lastly we can have one $W$ decay into $\tau\nu_\tau$ with the $\tau$ decaying
leptonically while the other $W$ decays into jets ($W \to qq'$). 
Again, we have one $b$ and $2j$ with 
one of the light quarks mistagged as a $\tau$-jet. 
\item[(iv)] 
$W + 2j$ processes: $pp \to W + 2j +X$ with the subsequent decays  
$W \to \ell\nu_\ell; \ell = e,\mu$ or $W \to \tau\nu_\tau$ with the 
$\tau$ decaying leptonically. Here, one jet is tagged or mistagged as 
a $b$ quark and the other mistagged as a $\tau$-jet.
\end{itemize}

Due to the huge cross-section for $pp \to q\bar{q}g$ with $q = b,c$, 
it is also pertinent to check that heavy quark 
semi-leptonic decays such as $b \to cl\nu$ do not overwhelm the signal. 
We find that this background is effectively cut to less than 
$10\%$ of the dominant background at all times by an isolation 
cut on the lepton $|\eta(\ell,j)| > 0.3$, the large rejection factor 
for non-$\tau$ jets, and the requirement $\notE_T > 20$ GeV.

For the lower integrated luminosity ($L$) of 30 fb$^{-1}$, 
we require $p_T(b,j) > 15$ GeV and $|\eta(b,j)| < 2.5$. 
The $b$-tagging efficiency ($\epsilon_b$) is taken to be $60\%$, 
the probability that a $c$-jet is mistagged as a $b$-jet ($\epsilon_c$)
is $10\%$ and 
the probability that any other jet is mistagged as a $b$-jet ($\epsilon_j$)
is taken to be $1\%$. For the higher luminosity $L = 300$ fb$^{-1}$, 
we take $\epsilon_b = 50\%$ and $p_T(b,j) > 30$ GeV \cite{ATLAS}.

%
%
In order to improve the signal significance we also apply a cut 
on the transverse mass of $m_T(\ell,\notE_T) < 30$ GeV. 
Using the definition of transverse mass given in \cite{barger} we find that 
this is very effective in controlling the $W+2j$ and
$t\bar{t}$ backgrounds.
In addition we require $\phi (\ell,\tau-{\rm jet}) < 170^o$, as
suggested by ATLAS and CMS collaborations \cite{ATLAS,CMS}, 
for the reconstruction of the Higgs mass as the
invariant mass of tau pairs.

We have applied a K factor of 1.3 for the DY background 
\cite{Campbell}, a K factor of 2 for $t\bar{t}$ \cite{ttbar}, 
a K factor of 1.5 for $tW$ \cite{Zhu}, 
a K factor of 0.9 for $W+2j$ \cite{w2j}, and a K factor of 2 
for $bq \to Wbq, q = u, d, s, c$ \cite{wbq} to include NLO effects. 
In order to further cut down the $t\bar{t}$ background, we apply a 
veto on events with more than 2 jets in addition to the $b$ and $\tau$ jets. 
This is very effective because, in $t\bar{t} + X$ production, 
nearly $50\%$ of events have at least one gluon from initial or final 
state radiation that passes $p_T > 15$ GeV and $|\eta| < 2.5$ \cite{ttbar}. 
Such events are then vetoed. 
We are also able to reduce contributions from 
top production where one $W \to jj$ decay occurs.

We have employed the programs MADGRAPH \cite{Madgraph}
and HELAS \cite{Helas} to evaluate matrix elements for both signal 
and background processes.

\section{The Discovery Potential at the LHC}

Based on the cuts defined above we show in Figure 3  the signal and 
background cross sections for an integrated luminosity 
$L = 30$ fb$^{-1}$ and $L = 300$ fb$^{-1}$. 
The signal is shown for 
$\tan\beta = 10$ and 50, with a common mass for scalar quarks, scalar 
leptons, gluino, and the $\mu$ parameter from the Higgs term in the 
superpotential, 
$m_{\tilde{q}} = m_{\tilde{g}} = m_{\tilde{\ell}} = \mu = 1$ TeV. 
All tagging efficiencies and K factors discussed above are included. 

 
\begin{figure}[htb]
\centering\leavevmode
\epsfxsize=6in
\epsfbox{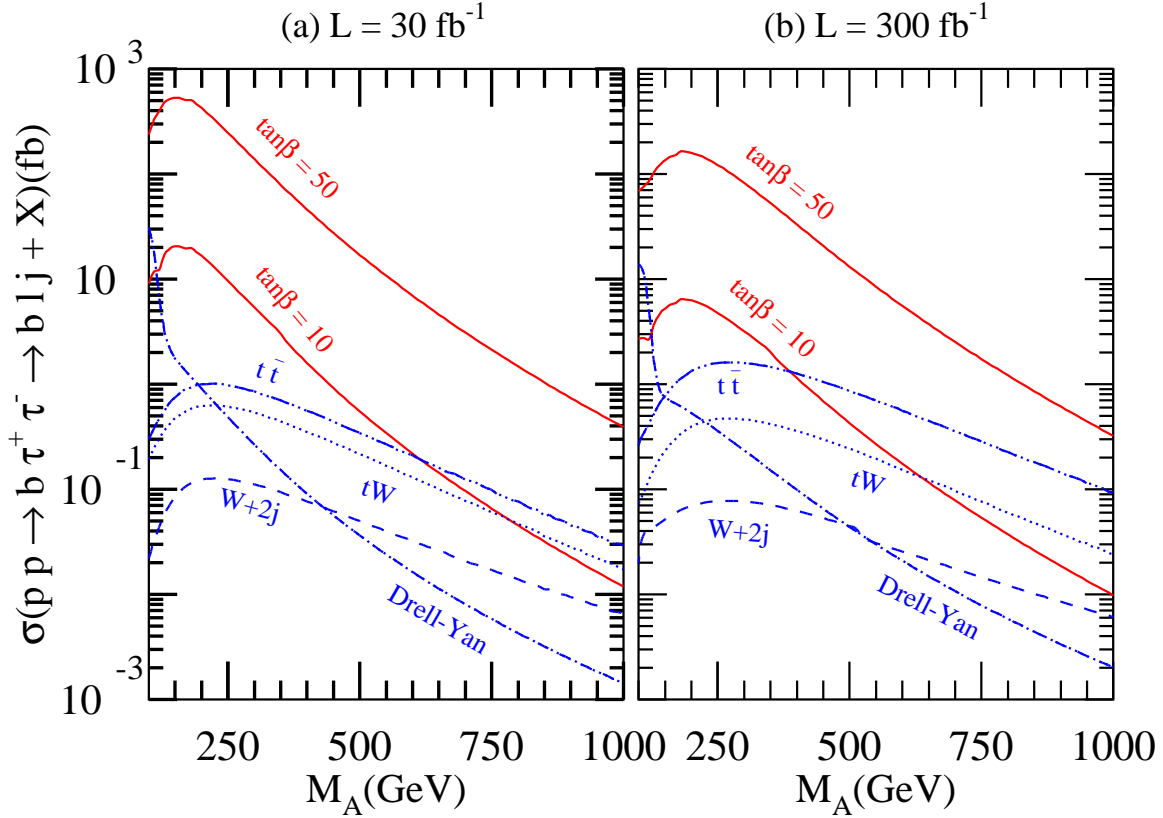}
\caption[]{
The signal cross section at the LHC for luminosity 
$L = 30$ fb$^{-1}$ and $300$ fb$^{-1}$,
as a function of $M_A$, and $\tan\beta = 10, 50$. 
Also shown are the background cross sections in the mass window of 
$M_A \pm \Delta M_{\tau\tau}$.
We have applied $K$ factors, acceptance cuts, and efficiencies of $b, \tau$ 
tagging and mistagging.} 
\label{fig:sigma}
%
\end{figure}

From this figure we note that the cross section of the Higgs signal 
with $\tan\beta \sim 50$ can be much larger than that of the physics 
background after acceptance cuts. The Drell-Yan processes make 
the major contributions to the physics background for Higgs mass 
$\alt 180$ GeV, but $t\bar{t}$ contributions become dominant for 
higher masses. The $W+2j$ contribution is very effectively controlled 
by the $b$ tagging requirement.

We define the signal to be observable 
if the lower limit on the signal plus background is larger than 
the corresponding upper limit on the background \cite{HGG,Brown}, namely,
\begin{eqnarray}
L (\sigma_s+\sigma_b) - N\sqrt{ L(\sigma_s+\sigma_b) } > 
L \sigma_b +N \sqrt{ L\sigma_b }\,\,,
\end{eqnarray}
which corresponds to
\begin{eqnarray}
\sigma_s > \frac{N^2}{L} \left[ 1+2\sqrt{L\sigma_b}/N \right]\,\,.
\end{eqnarray}
Here $L$ is the integrated luminosity, 
$\sigma_s$ is the cross section of the Higgs signal, 
and $\sigma_b$ is the background cross section.  
Both cross sections are taken to be 
within a bin of width $\pm\Delta M_{\tau\tau}$ centered at $M_\phi$. 
In this convention, $N = 2.5$  corresponds to a 5$\sigma$ signal.
We take the integrated luminosity $L$ to be 30 fb$^{-1}$ 
and 300 fb$^{-1}$ \cite{ATLAS}. 

For $\tan\beta \agt 10$, 
$M_A$ and $M_H$ are almost degenerate when $M_A \agt$ 125 GeV, 
while $M_A$ and $M_h$ are very close to each other for $M_A \alt$ 125 GeV.
Therefore, when computing the discovery reach, we add the cross sections 
of the $A^0$ and the $h^0$ for $M_A < 125$ GeV 
and those of the $A^0$ and the $H^0$ for $M_A \ge 125$ GeV 
\cite{Higgsmass}.

Figure 4 shows the 5$\sigma$ discovery contours for the MSSM Higgs bosons 
where the discovery region is the part of the parameter space above the 
curves. We have chosen 
$M_{\rm SUSY} = m_{\tilde{q}} = m_{\tilde{g}} = m_{\tilde{\ell}} 
= \mu = 1$ TeV.
If $M_{\rm SUSY}$ is smaller, the discovery region of 
$A^0,H^0 \to \tau^+\tau^-$ 
will be slightly reduced for $M_A \agt 250$ GeV,
because  the Higgs bosons can decay into SUSY particles \cite{HZ2Z2} 
and the branching fraction of $\phi \to \tau^+\tau^-$ is suppressed.
For $M_A \alt 125$ GeV, the discovery region of $H^0 \to \tau^+\tau^-$ 
is slightly enlarged for a smaller $M_{\rm SUSY}$, 
but the observable region of $h^0 \to \tau^+\tau^-$ is slightly reduced 
because the lighter top squarks make the $H^0$ and the $h^0$ lighter; 
also the $H^0 b\bar{b}$ coupling is enhanced 
while the $h^0 b\bar{b}$ coupling is reduced \cite{Higgsmass}.

 
\begin{figure}[htb]
\centering\leavevmode
\epsfxsize=5in
\epsfbox{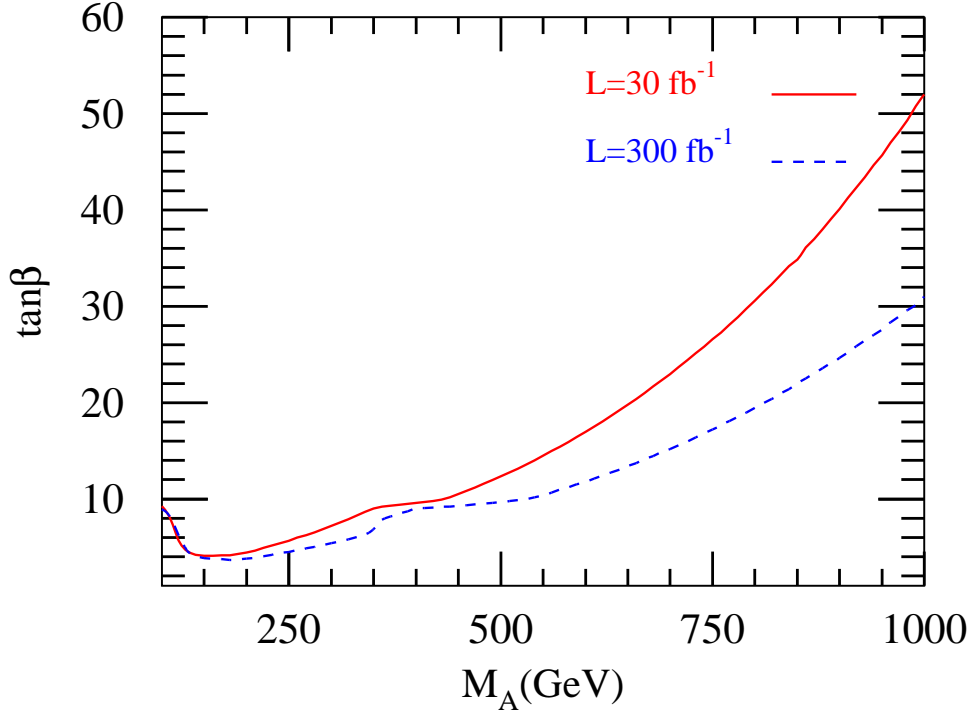}
\caption[]{
The $5\sigma$ discovery contours at the LHC 
for an integrated luminosity ($L$) of 30 fb$^{-1}$, and 
300 fb$^{-1}$ in the $(M_A,\tan\beta)$ plane.  
The signal includes $\phi = A^0$ and $h^0$ for $M_A < 125$ GeV, 
and $\phi = A^0$ and $H^0$ for $M_A \ge 125$ GeV except that, for 
$\tan\beta < 10$, $\phi = A^0$ only. The discovery region is the 
part of the parameter space above the contours.}
\label{fig:contour-atlas}
\end{figure}

We find that the discovery contour even dips below $\tan\beta = 10$ for 
$100$ GeV $< M_A < 300 - 400$ GeV depending on luminosity. 
Below $\tan\beta = 10$ our approximation of mass degeneracy of MSSM 
Higgs bosons breaks down; therefore we include only one Higgs boson
$(A^0)$ in our calculations.

\section{The Discovery Potential at the Fermilab Tevatron}

To study the discovery potential of this channel at the Fermilab
Tevatron Run II, we require
\begin{itemize}
\item
one $b$ quark with $p_T(b) > 15$ GeV, $|\eta(b)| < 2.5$ and a tagging
efficiency $\eps_b = 60\%$,
\item
one isolated lepton with $p_T(\ell) > 10$ GeV and $|\eta(\ell)| < 2.0$,
\item
one jet with $p_T(j) > 15$ GeV and $|\eta(j)| < 2.5$ for the tau jet, 
and a tagging efficiency of $38\%$,
\item
the transverse missing energy ($\notE_T$) should be greater than 20 GeV,
\item
the transverse mass of the lepton and missing transverse energy, 
$M_T(\ell,\notE_T)$, should be less than 30 GeV,
\item
the transverse angular separation of the lepton and tau jet,
$\phi(\ell,j)$, should be less than $170^o$, 
\item
the energy fractions for the lepton and the tau jet should be 
between 0 and 1, $(0 \le x_\ell, x_h \le 1)$, and 
\item
the invariant mass of the reconstructed tau pairs should be within the
mass window of the Higg mass with $\Delta M_{\tau\tau} = 0.15 M_\phi$
\end{itemize}

In Figure 5 we show the signal and background cross sections for the 
Fermilab Tevatron. The signal is shown for 
$\tan\beta = 10$ and 50, with a common mass for scalar quarks, scalar 
leptons and the gluino $m_{\tilde{q}} = m_{\tilde{g}} 
= m_{\tilde{\ell}} = \mu = 1$ TeV. 
All tagging efficiencies and K factors discussed above are included. 

 
\begin{figure}[htb]
\centering\leavevmode
\epsfxsize=5in
\epsfbox{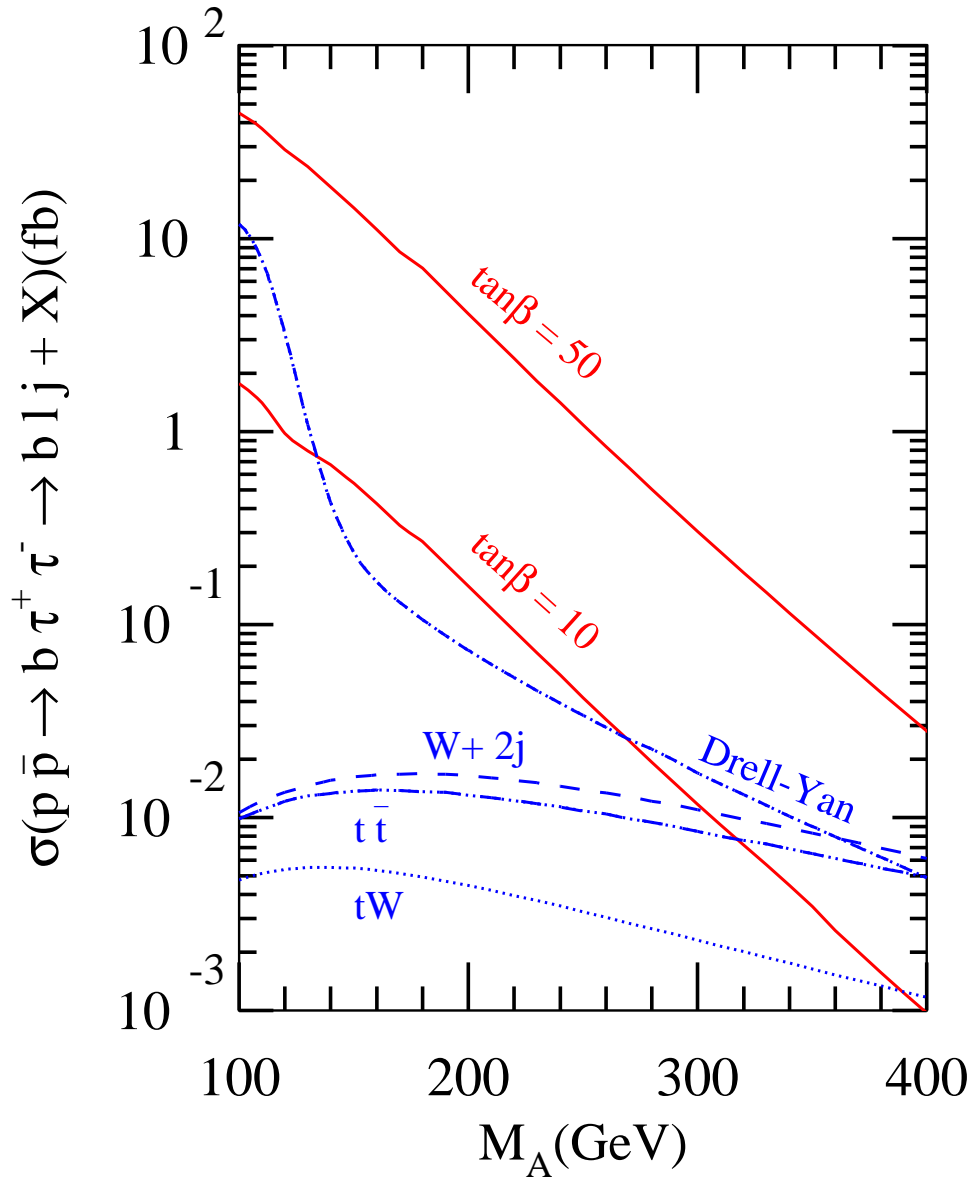}
\caption[]{
The signal cross section at the Fermilab Tevatron Run II as a function
of $M_A$, and $\tan\beta = 10, 50$. 
Also shown are the background cross sections in the mass window of 
$M_A \pm \Delta M_{\tau\tau}$.
We have applied $K$ factors, acceptance cuts, and efficiencies of $b, \tau$ 
tagging and mistagging. 
\label{fig:sigma-cdf}
}
\end{figure}
From this figure we note that while $b\tau\tau$ and $t\bar{t}$ 
make major contributions to the physics background at the LHC, 
$b\tau\tau$ and $Wjj$ become the dominant background at the Tevatron 
for $M_A < 400$ GeV. 
The cross section of the Higgs signal 
with $\tan\beta \sim 50$ can be much larger than that of the physics 
background after acceptance cuts.

Figure 6 shows the 5$\sigma$ discovery contours for the MSSM Higgs bosons 
where the discovery region is the part of the parameter space above the 
curves. 

 
\begin{figure}[htb]
\centering\leavevmode
\epsfxsize=5in
\epsfbox{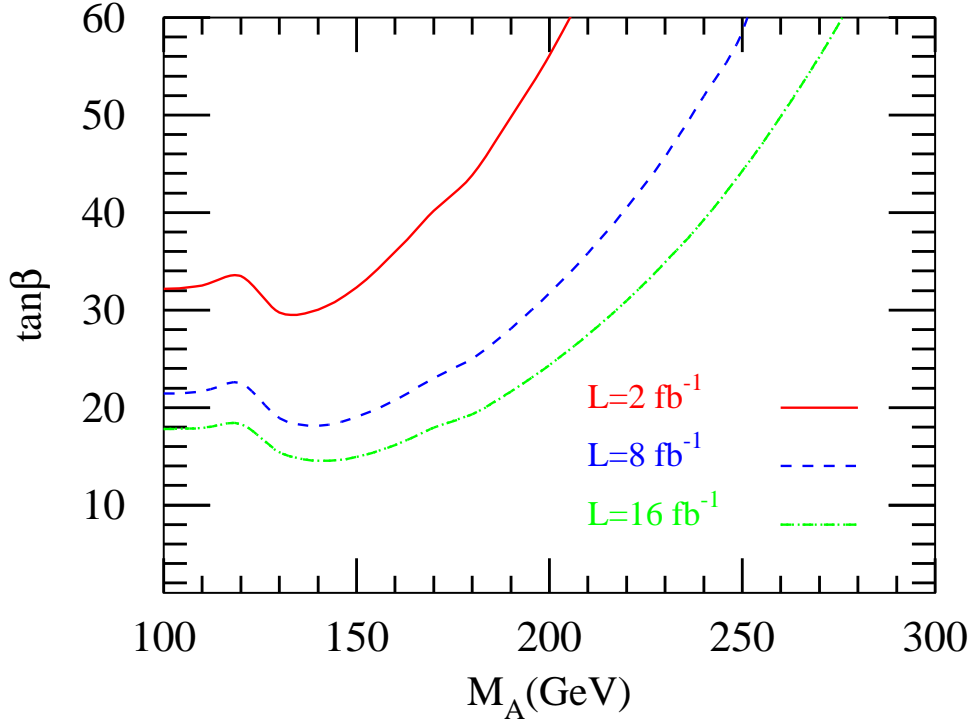}
\caption[]{
The $5\sigma$ discovery contours at the Fermilab Tevatron Run II
for an integrated luminosity ($L$) of 2 fb$^{-1}$, 8 fb$^{-1}$, 
16 fb$^{-1}$ in the $M_A$ versus $\tan\beta$ plane.  
The signal includes $\phi = A^0$ and $h^0$ for $M_A < 125$ GeV, 
and $\phi = A^0$ and $H^0$ for $M_A \ge 125$ GeV except for 
$\tan\beta < 10$ where $\phi = A^0$ only. The discovery region is the 
part of the parameter space above the contours.}
\label{fig:contour-cdf}
\end{figure}
We find that the discovery contour for the Tevatron Run II can be
slightly below $\tan\beta = 30$ with an integrated luminosity 
of 2 fb$^{-1}$ and below $\tan\beta = 20$ with $L \simeq 8$ fb$^{-1}$
for $M_A < 150$.
For $\tan\beta \sim 50$, the Tevatron Run II will be able to discovery
the Higgs bosons up to $M_A \sim 200$ GeV with $L =$ 2 fb$^{-1}$
and up to $M_A \sim 250$ GeV with $L \sim$ 8 fb$^{-1}$.

\section{Conclusions}

The tau pair decay mode is a promising channel for the discovery of 
the neutral Higgs bosons in the minimal supersymmetric model at the LHC. 
The $A^0$ and the $H^0$ should be observable in a large region 
of parameter space with $\tan\beta \agt 10$.
In particular, Fig. 4 shows that the associated final state of 
$b\phi \to b\tau^+\tau^-$ 
could discover the $A^0$ and the $H^0$ at the LHC 
with an integrated luminosity of 30 fb$^{-1}$ if $M_A \alt 800$ GeV.
At a higher luminosity of 300 fb$^{-1}$, the discovery region in 
$M_A$ is easily expanded up to $M_A = 1$ TeV for $\tan\beta \sim 50$.

%
%
In Figure 7, we compare the LHC discovery potential of $b\phi^0$
production for the muon pair discovery channel, as determined 
in Ref.\cite{hbmm}, and the tau pair discovery channel, 
for an integrated luminosity of 30 fb$^{-1}$. 
It is clear that the tau pair channel can be discovered
in a larger region of the parameter space. However, the muon pair
channel can also be observable in a significantly large region. In
addition, the muon pair channel will provide a good opportunity to 
precisely reconstruct the masses for MSSM Higgs bosons.
The discovery of the associated final states of 
$b\phi \to b\tau^+\tau^-$ and $b\phi \to b\mu^+\mu^-$ 
will provide information about the Yukawa couplings of $b\bar{b}\phi$ 
and an opportunity to measure $\tan\beta$. 
The discovery of both $\phi \to \tau^+\tau^-$ and $\phi \to \mu^+\mu^-$ 
will allow us to study the Higgs Yukawa couplings with the leptons.

 
\begin{figure}[htb]
\centering\leavevmode
\epsfxsize=5in
\epsfbox{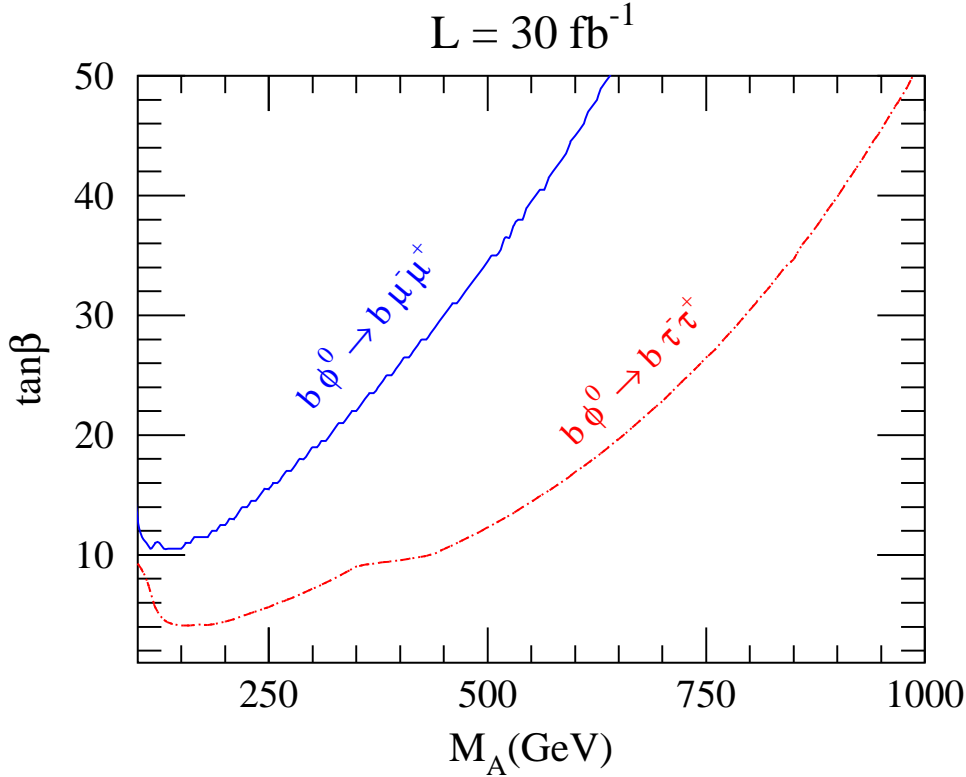}
\caption[]{
The $5\sigma$ discovery contours at the LHC
for $b\phi^0 \to b\mu^-\mu^+$ and $b\phi^0 \to b\tau^-\tau^+$ 
with an integrated luminosity ($L$) of 30 fb$^{-1}$.
}
\label{fig:contour-hbll}
\end{figure}

We find that the discovery contour for the Tevatron Run II can be
slightly below $\tan\beta = 30$ with an integrated luminosity ($L$) 
of 2 fb$^{-1}$ and below $\tan\beta = 20$ with $L \simeq 8$ fb$^{-1}$
for $M_A < 150$.
For $\tan\beta \sim 50$, the Tevatron Run II will be able to discover
the Higgs bosons of MSSM up to $M_A \sim 200$ GeV with $L =$ 2 fb$^{-1}$
and up to $M_A \sim 250$ GeV with $L \sim$ 8 fb$^{-1}$. 

%
%
The inclusive tau pair channel ($\phi^0 \to \tau^-\tau^+$) has been
studied by the ATLAS \cite{Richter-Was,ATLAS} and the CMS \cite{CMS} 
collaborations with realistic simulations. 
Both collaborations have confirmed that this channel will offer great 
promise at the LHC. Our results for $b\phi^0 \to b\tau^-\tau^+$ 
are consistent with those given in these references and also 
with the results of Ref.~\cite{Carena:2006ai} for the Fermilab Tevatron.

\section*{Acknowledgments}

We are grateful to David Rainwater for beneficial discussions. 
C.K. thanks the Stanford Linear Accelerator Center 
for hospitality and support during a sabbatical visit. 
R.M. is grateful to John Campbell for his help in using the program MCFM 
for NLO calculations of the $bH$ process. 
Part of our computing resources were provided by 
the OU Supercomputing Center for Education and Research (OSCER) 
at the University of Oklahoma.
This research was supported 
in part by the U.S. Department of Energy
under Grants 
No.~DE-AC02-98CH10886,
No.~DE-FG03-98ER41066, 
No.~DE-FG02-03ER46040,
No.~DE-FG03-93ER40757, and
No.~DE-AC02-76SF00515.

\newpage


\end{document}